\documentclass[conference]{IEEEtran}
\IEEEoverridecommandlockouts
\usepackage{cite}
\usepackage{amsmath,amssymb,amsfonts}
\usepackage{algorithmic}
\usepackage{algorithm}
\usepackage{subfigure}
\usepackage{graphicx}
\usepackage{textcomp}
\usepackage{xcolor}
\usepackage{booktabs}
\usepackage{multirow}
\usepackage{makecell}

\def\BibTeX{{\rm B\kern-.05em{\sc i\kern-.025em b}\kern-.08em
    T\kern-.1667em\lower.7ex\hbox{E}\kern-.125emX}}
\begin{document}

\title{Efficient Training Approaches for Performance Anomaly Detection Models in Edge Computing Environments
}

\author{Duneesha Fernando, 
Maria A. Rodriguez, 
Patricia Arroba, 
Leila Ismail, 
Rajkumar Buyya 
\thanks{D. Fernando, M. A. Rodriguez and R. Buyya are with the Cloud Computing and Distributed Systems (CLOUDS) Laboratory, School of Computing and Information
Systems, The University of Melbourne, Parkville, Australia.}
\thanks{P. Arroba is with the CCS–Center for Computational Simulation, Universidad Politécnica de Madrid, Madrid, Spain.}
\thanks{L. Ismail is with the Intelligent Distributed Computing and Systems Research (INDUCE) Lab, Department of Computer Science and Software Engineering, United Arab Emirates University, United Arab Emirates.}
}

\maketitle
\thispagestyle{plain}
\pagestyle{plain}

\begin{abstract}
Microservice architectures are increasingly used to modularize IoT applications and deploy them in distributed and heterogeneous edge computing environments. Over time, these microservice-based IoT applications are susceptible to performance anomalies caused by resource hogging (e.g., CPU or memory), resource contention, etc., which can negatively impact their Quality of Service and violate their Service Level Agreements. Existing research on performance anomaly detection for edge computing environments focuses on model training approaches that either achieve high accuracy at the expense of a time-consuming and resource-intensive training process or prioritize training efficiency at the cost of lower accuracy. To address this gap, while considering the resource constraints and the large number of devices in modern edge platforms, 
we propose two clustering-based model training approaches : (1) intra-cluster parameter transfer learning-based model training (ICPTL) and (2) cluster-level model training (CM). These approaches aim to find a trade-off between the training efficiency of anomaly detection models and their accuracy. We compared the models trained under ICPTL and CM to models trained for specific devices (most accurate, least efficient) and a single general model trained for all devices (least accurate, most efficient). Our findings show that ICPTL’s model accuracy is comparable to that of the model per device approach while requiring only 40\% of the training time. In addition, CM further improves training efficiency by requiring 23\% less training time and reducing the number of trained models by approximately 66\% compared to ICPTL, yet achieving a higher accuracy than a single general model.
\end{abstract}


\section{Introduction}
\label{sec:introduction}

The emergence of the Internet of Things (IoT) paradigm, supporting a wide range of smart services for application domains such as healthcare, transportation, industrialization, and agriculture, has led to an exponential rise in the number of IoT devices globally. Initially, the enormous quantities of data generated by IoT devices were sent to the cloud for processing, giving an upsurge to cloud-centric IoT. However, data transmission towards centralized cloud data centers increases network congestion and latencies. To overcome this limitation, edge computing, where processing is performed closer to the IoT devices, was introduced. Edge platforms consist of a wide range of devices (e.g., smart routers and switches, edge servers, micro-datacentres, etc.) that are heterogeneous in terms of computing, storage, and networking capabilities. 

\begin{figure}[t]
\centerline{\includegraphics[width=0.5\textwidth]{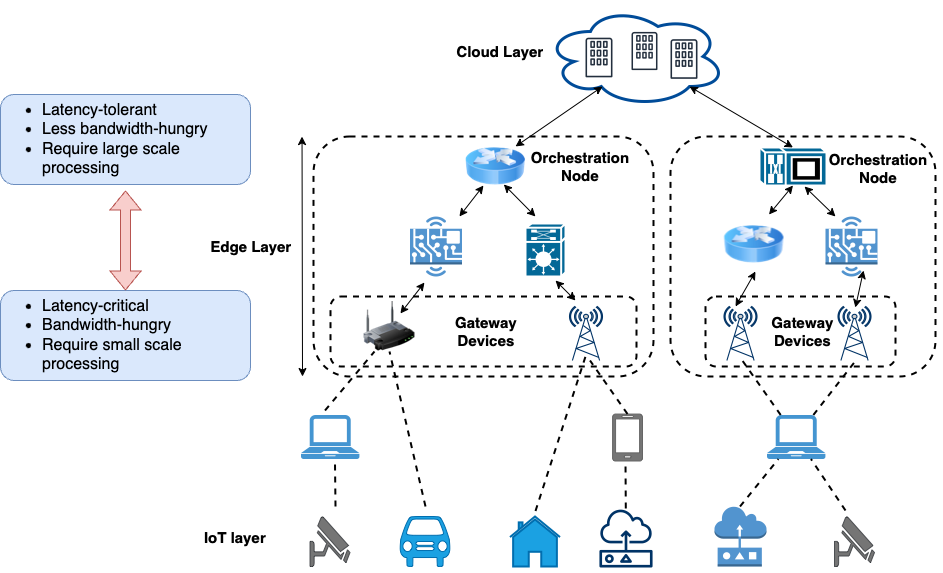}}
\caption{Properties of applications placed in edge computing environments}
\label{microservice_placement}
\vspace{-15pt}
\end{figure}

The computing and storage capacities of edge devices are higher than that of IoT devices but lower than that of the cloud. Additionally, the computing and storage capabilities of edge devices are arranged such that the edge devices closer to the IoT devices have lower computing and storage capabilities than the edge devices closer to the cloud \cite{pallewatta2023placement, Lee2018hierarchical}. As a result, edge and cloud resources are used dynamically to place IoT applications based on their Quality of Service (QoS) and resource requirements. Therefore, as shown in Figure \ref{microservice_placement}, IoT applications that are latency-critical, bandwidth-hungry, and require small-scale processing are deployed in edge devices closer to IoT devices. In contrast, latency-tolerant IoT applications that require large-scale processing are deployed at edge devices closer to the cloud or even at cloud data centers \cite{Golpayegani2024Adaptation, li2021fuzzy}. This paper refers to such edge-cloud integrated environments as “edge computing environments”.

In order to be deployed in distributed and heterogeneous edge computing environments, IoT applications are modeled as interdependent, lightweight, and granular modules. Microservice architectures that build distributed applications based on loosely coupled, lightweight, independently deployable, and scalable modules are increasingly being used for this purpose. Several studies have utilized the abstraction provided by microservice architectures to model IoT applications \cite{al-doghman2023ai-enabled, WU2023Towards} and have developed algorithms to schedule those containerized microservice-based IoT applications in edge computing environments \cite{pallewataa2022qos, shirvani2023multiobjective}. This granular placement ensures that the QoS requirements of each module can be satisfied while maximizing resource usage.

However, long-running microservice-based IoT applications, such as smart health monitoring and smart road traffic management applications, deployed in edge platforms may be prone to a degradation in performance or performance anomalies, mainly due to the environment's highly dynamic and multi-tenant nature \cite{becker2020towards, Soualhia2019infrastructure}. For instance, an upsurge in workload may result in an application hogging a resource (e.g., CPU or memory), or the co-location of microservices may result in resource contention, both cases potentially negatively impacting the performance of a given application. Furthermore, several types of malicious attacks, such as distributed denial-of-service (DDoS) attacks, also indirectly cause performance problems \cite{hunter2022deep}. These performance anomalies occurring in edge computing environments adversely affect the QoS of microservice-based IoT applications and lead to violations of their Service Level Agreements (SLAs). Therefore, it is crucial to continuously monitor microservice-based IoT applications in edge computing environments with the aim of detecting and eventually mitigating such anomalies. This paper specifically focuses on anomaly detection, which is the initial step in ensuring the smooth operation of edge computing environments.

Existing research studies on performance anomaly detection for edge computing environments advocate for using unsupervised multivariate time-series methods \cite{becker2020towards, Soualhia2019infrastructure, tuli2021generative, Skaperas2024pragmatical}, where a model learns the normal performance data distribution of microservices and identifies deviations as anomalies. Microservices deployed across devices in an edge infrastructure have heterogeneous QoS and resource requirements, resulting in heterogeneous normal data distributions. One approach to machine learning-based anomaly detection at the edge is to train a single, generic anomaly detection model (GM) using data from all microservices across all devices \cite{Soualhia2019infrastructure, peltonen2023linkedge, Psaromanolakis2023MLOps}. Training a single model is efficient in terms of training time and resource usage, and reduces the complexity associated with model management. However, it requires aggregating local (edge device) time series data in a centralized location, usually a cloud data center, thus considerably increasing network utilization and potentially causing network congestion \cite{tuli2021generative}. Importantly, a generic model usually yields lower accuracy due to the difficulty of generalizing across the diverse normal data distributions. An alternative approach involves training a ``model per edge device" (MPD) using data collected from the microservices deployed in that specific edge device \cite{becker2020towards,tuli2021generative,john202ai, raj2021edge_mlops}. Since microservices within a single edge device usually have similar QoS and resource requirements \cite{pallewataa2022qos, fu2021qos} and hence have similar normal data distributions, this approach results in models with higher accuracy, as the model is expected to specialize well across similar normal data distributions. Training as many models as edge devices consumes a significant amount of resources and results in longer aggregate training times, as well as increased model management complexity. While a model per device allows for each edge device to train its associated anomaly detection model, performing training, which is a resource-intensive process, might be challenging for most edge devices due to their resource-constrained nature. 

The aforementioned approaches (which we identify as baselines) represent extremes in terms of accuracy and efficiency; a GM has low accuracy but can be efficiently trained, while MPDs yield high accuracy, but the training process is time-consuming and resource-hungry. Considering the resource constraints and large number of devices present in modern edge platforms, this research aims to bridge the gap between GM and MPD by finding a trade-off between the training efficiency of anomaly detection models and their accuracy. We propose to achieve this by conducting model training within clusters of edge devices with similar normal data distributions. To that end, we propose a similarity metric capable of clustering edge devices with similar normal data distributions. Furthermore, we introduce two clustering-based training approaches, namely, (1) intra-cluster parameter transfer learning-based model training (ICPTL) and (2) cluster-level model training (CM). ICPTL aims to match MPD's accuracy with fewer training cycles, while CM aims to further improve training efficiency by reducing the number of models. To the best of our knowledge, this is the first work to propose anomaly detection model training approaches tailored for the specific characteristics and constraints of edge environments with the aim of increasing training efficiency while achieving high model accuracy. 

We conducted a comprehensive and reproducible evaluation of the proposed clustering-based training approaches using a publicly available and widely-used performance anomaly dataset called the ``Server Machine Dataset (SMD)'' \cite{su2019robust}, which consists of devices with heterogeneous normal data distributions. Initially, we compared four common unsupervised multivariate time-series anomaly detection techniques for accuracy and efficiency (in terms of resource utilization and detection time) against the SMD dataset to determine the most suitable algorithm for the rest of our evaluations. Upon evaluating the proposed clustering-based training approaches using the identified algorithm, we demonstrate that the clustering approaches achieve a balance between accuracy and efficiency, as expected. To further validate the results, we conducted a small-scale experiment using data collected from a set of microservices with heterogeneous QoS and resource requirements deployed in an emulated edge computing environment \cite{negin2024icontinuum}.

The rest of the paper is organized as follows: Section \ref{sec:related work} reviews the existing related works. Section \ref{sec:main_approach} presents the clustering-based training approaches. Section \ref{sec:eval} evaluates the performance anomaly detection algorithms and the proposed training approaches. Section \ref{sec:dynamism} discusses how the proposed training approaches handle the dynamic nature of edge computing environments. Section \ref{sec:conclusion} concludes the paper and draws future research directions.

\section{Related Work}
\label{sec:related work}

\begin{table*}[t]
\caption{A summary of performance anomaly detection studies in edge computing environments}
\label{table:relatedWork}
\renewcommand{\arraystretch}{1.5}
\resizebox{\textwidth}{!}{%
\begin{tabular}{|c|c|c|c|c|c|}
\hline
\multirow{2}{*}{\textbf{Work}} &
  \multirow{2}{*}{\textbf{Anomaly Detection Algorithms}} &
  \multicolumn{2}{c|}{\textbf{Anomaly Detection Evaluation Aspects}} &
  \multirow{2}{*}{\textbf{Model Training Approach}} &
  \multirow{2}{*}{\textbf{Main Objective of Work}}\\
  \cline{3-4}
  & & \textbf{Accuracy} & \textbf{Efficiency} & & \\
  \hline
  \cite{Soualhia2019infrastructure} &
   \makecell{ARIMA(Conventional),MC(ML), SVM(ML),\\ RF(ML), NN(ML), BN(ML), TAN(ML)} & \checkmark & $\times$ & Generic Model (GM) &
  \makecell{Framework that performs anomaly detection\\ in edge environments.}\\ 
  \hline
  \cite{becker2020towards} & \makecell{ARIMA(Conventional), BIRCH(ML), \\ LSTM(DNN), EMA(Conventional)} & $\times$ & \makecell{\checkmark CPU utilization, \\ detection delay} & Model per device (MPD) & \makecell{AIOPS framework for edge environments.}\\
  \hline
  \cite{tuli2021generative} & \makecell{OCSVM(ML), IF(ML), DILOF (DL), \\ ONLAD(ML), CAE-M(DL), USAD(DL),\\  MAD-GAN(DL), SlimGAN(DL)} & \checkmark & \makecell{\checkmark memory consumption, \\ inference time} & Model per device (MPD) & \makecell{Propose memory efficient anomaly \\ detection approaches for edge devices.}\\
  \hline
  \cite{Skaperas2024pragmatical} & \makecell{CPD approaches : Bayesian (Conventional), \\ CUSUM (Conventional)} & \checkmark & \makecell{\checkmark resource utilization \\ (CPU and memory consumption), \\ detection delay} & Model per device (MPD) & \makecell{Framework that facilitates evaluation of\\ CPD algorithms.}\\
  \hline
  Our work & \makecell{VAR(Conventional), IF(ML), AE(DNN), \\ LSTM-AE(DNN)} & \checkmark & \makecell{\checkmark detection time, resource \\requirements (CPU and memory)} & \makecell{Similarity-based clustering\\ approaches} & \makecell{Propose efficient model training  approaches \\ for performance anomaly detection in \\ edge environments.}\\
  \hline
  \multicolumn{6}{|l|}{\begin{tabular}{l}BIRCH: Balanced Iterative Reducing and Clustering using Hierarchies, EMA: Exponential Moving Average, LSTM: Long Short-Term Memory, SVM: Support Vector Machine, RF: Random Forest, \\NN: Neural Network, MC: Markov Chain, BN: Bayesian Network, IF: Isolation Forest, TAN: Tree-Augmented Naive Bayes, OCSVM: One-Class Support Vector Machine, DILOF: Deep Isolation \\Forest, ONLAD: Online Anomaly Detection via Adaptive Learning, CAE-M: Convolutional Autoencoder-Mahalanobis, USAD: Unsupervised Anomaly Detection, MAD-GAN: Multi-scale Anomaly\\ Detection with GANs, SlimGAN: Slim Generative Adversarial Network, CUSUM: Cumulative Sum Control Chart, VAR: Vector AutoRegression, AE: AutoEncoder, LSTM-AE: LSTM-AutoEncoder\end{tabular}} \\ \hline
\end{tabular}%
}
\end{table*}

In this section, we provide an overview of performance anomaly detection studies conducted in edge computing environments in terms of the algorithms, evaluation metrics, and training methods they have used. We also explore several related works on machine learning (ML) model training approaches employed in edge computing environments.

\subsection{Performance anomaly detection in edge computing environments}
\label{subsec:perf_ad_edge}

Several works have proposed monitoring solutions for edge computing environments, some of which have incorporated alerting as a functionality \cite{taherizadeh2018monitoring, prasad2019influence}. However, none of these monitoring solutions perform any advanced form of anomaly detection other than threshold-based alerting. Threshold-based alerting requires system administrators to manually define thresholds for each metric per device, which is not scalable given the large number of edge devices and the multitude of performance metrics being monitored. Moreover, these static thresholds become obsolete faster in dynamic edge computing environments. In contrast, multivariate time series anomaly detection algorithms can accurately detect more types of anomalies since they consider inter-metric correlations.

Building on the limitations of threshold-based alerting, recent research on performance anomaly detection for edge computing environments suggests using unsupervised multivariate time-series techniques for anomaly detection \cite{becker2020towards, Soualhia2019infrastructure, tuli2021generative, Skaperas2024pragmatical}. Becker et al. evaluated ARIMA, BIRCH, LSTM, and EMA algorithms, all of which are unsupervised anomaly detection approaches \cite{becker2020towards}. Similarly, Skaperas et al. evaluated two change point (CP) detection approaches: Bayesian and cumulative sum (CUSUM), both of which can be identified as unsupervised approaches \cite{Skaperas2024pragmatical}. Tuli et al. explored several unsupervised approaches, including OCSVM, IF, DILOF, and USAD \cite{tuli2021generative}. Although Soualhia et al. evaluated a group of supervised approaches, they emphasized the infeasibility of such methods due to the need for labeled normal and anomaly data for training \cite{Soualhia2019infrastructure}. As anomaly data may not be available during the initial execution of applications and labeling can be costly, they suggest using unsupervised anomaly detection approaches. These algorithms are trained on unlabeled normal data to learn the normal data distribution and identify deviations from it as anomalies. Therefore, in our work, we utilize unsupervised multivariate time-series anomaly detection techniques.

Although research on performance anomaly detection in edge computing environments is still emerging, AI-based performance anomaly detection in cloud environments is well-established \cite{kardani2019performance, audibert2020usad, naikade2020automated, audibert2022dodeep}. It is evident that even the research on performance anomaly detection for edge environments leverages cloud anomaly detection approaches \cite{becker2020towards, Soualhia2019infrastructure, Skaperas2024pragmatical, tuli2021generative}. According to Audibert et al. \cite{audibert2022dodeep}, existing anomaly detection algorithms can be broadly classified into three categories: (1) conventional methods, (2) machine learning (ML)-based methods, and (3) deep neural network (DNN)-based methods. Therefore, in section \ref{subsec:algo_eval}, where we evaluate common unsupervised multivariate time-series anomaly detection techniques against the SMD dataset to identify the most suitable algorithm for the rest of our evaluations, we select four representative algorithms from each of these categories. 

Cloud performance anomaly detection research typically does not account for the constraints of edge computing environments and, therefore, only focuses on evaluating the accuracy of anomaly detection models. However, due to the resource-constrained nature of edge devices, edge performance anomaly detection studies evaluate both accuracy and efficiency during inference (in terms of resource utilization and detection time) when identifying a suitable performance anomaly detection algorithm to be deployed at the edge devices \cite{becker2020towards, tuli2021generative, Skaperas2024pragmatical}. Consequently, in section \ref{subsec:algo_eval}, we evaluate the accuracy as well as resource utilization efficiency of four representative algorithms from each category of Audibert's taxonomy to determine the most suitable algorithm for the remainder of our evaluations.

In addition to evaluating the accuracy and efficiency of the algorithms during detection, considering the fact that these models are trained on resource-constrained edge devices, some of these studies have also assessed the time taken and resource utilization during the training process \cite{tuli2021generative, Soualhia2019infrastructure}. Although Becker et al. do not evaluate the above aspect, they also propose training a model at the edge device \cite{becker2020towards}. As discussed in section \ref{sec:introduction}, while this MPD approach yields high accuracy, the presence of a large number of edge devices leads to a high aggregate training time and an increased total consumption of resources during the training process. On the other hand, Soualhia et al. propose training a generic anomaly detection model (GM) using data from all devices \cite{Soualhia2019infrastructure}. While this approach can be efficiently trained, it results in low accuracy. A comparison between prior works on performance anomaly detection in edge computing environments \cite{Soualhia2019infrastructure, becker2020towards, tuli2021generative, Skaperas2024pragmatical} along with our proposed work is shown in Table \ref{table:relatedWork}. Notably, none of these studies have tried to reach a tradeoff between the
efficiency and accuracy of performance anomaly detection model training, which presents a gap in current research that we aim to explore further in this paper.

\subsection{Machine learning model training approaches in edge computing environments}
\label{subsec:eff_train}

Since none of the edge performance anomaly detection studies have developed approaches that strike a balance between the efficiency and accuracy of performance anomaly detection model training, we reviewed the literature on ML model training for Artificial Intelligence of Things (AIoT) applications as well \cite{john202ai,raj2021edge_mlops, peltonen2023linkedge, Psaromanolakis2023MLOps}. Our goal was to identify the model training approaches used in these works. John et al. \cite{john202ai} and Raj et al. \cite{raj2021edge_mlops} discussed the generic and specialized model training approaches that we could also identify from edge performance anomaly detection literature, while Peltonen et al. \cite{peltonen2023linkedge} and Psaromanolakis et al. \cite{Psaromanolakis2023MLOps} focused on training a generic model. As mentioned before, since these two training approaches achieve either high accuracy or high efficiency, we use them as the baselines for comparing our proposed clustering-based training approaches.

Additionally, a few works employ novel ML approaches, such as federated learning \cite{schneible2017anomaly} and distributed ML \cite{yang2020fog}, for model training in edge computing environments. Schneible et al. utilized federated learning in the context of supervised anomaly detection \cite{schneible2017anomaly}. Since different edge devices encounter different anomaly data, they distribute the required updates via a centralized model to train the anomaly detection models of other edge devices with newly seen anomaly data. However, adopting this approach for unsupervised performance anomaly detection would result in each model learning the normal data distributions of all microservices deployed across all edge devices, ultimately leading to lower model accuracy. 

On the other hand, Yang et al. \cite{yang2020fog} used distributed ML to detect network anomalies at edge devices by analyzing network packets. Their approach involves centrally training an ML model using multiple interconnected computing resources and then deploying it to infer anomalies at the edge device level. Although this approach makes the training process faster by parallelizing and utilizing more resources, it results in a generic model trained on data aggregated from all edge devices. This can lead to reduced accuracy when applied to performance anomaly detection model training, as the generic model is not capable of generalizing well across the heterogeneous normal data distributions of all microservices.

Thus, adapting ML approaches such as federated learning and distributed ML for training performance anomaly detection models results in the creation of a generic model where every model learns the normal data distribution of microservices deployed in other devices as well. Although this approach results in high accuracy during situations where all the collected data belong to the same independent and identically distributed (IID) distribution, in our case, since the data collected from different microservices belong to non-IIDs, this approach could drift the model away from expected behavior, resulting in low accuracy. Therefore, it is clear that simply adapting existing ML approaches to perform anomaly detection model training in edge computing environments does not allow us to strike a balance between efficiency and accuracy. Therefore, in this paper, we propose a new approach to training anomaly detection models tailored to the specific characteristics and constraints of edge environments with the aim of increasing training efficiency while achieving high model accuracy.

\section{Clustering-based Training Approaches}
\label{sec:main_approach}

This study explores a hierarchical edge environment with multiple heterogeneous devices \cite{pallewatta2019microservices}. If there are $N$ edge devices, the set of edge devices can be defined as $E$, where $e_i$ represents the $i^{th}$ edge device. 
\begin{equation}
E = \{e_1,e_2,...,e_N\}
\end{equation}
Each device is equipped with a monitoring agent that gathers $M$ performance and resource consumption metrics from each microservice deployed on that device. Consistent with prior research on performance anomaly detection at the edge, this study also presumes that an unsupervised multivariate anomaly detection model, which analyzes $M$ performance metrics, is deployed at each edge device \cite{becker2020towards, Soualhia2019infrastructure, tuli2021generative, Psaromanolakis2023MLOps}. Normal performance data (i.e., data collected under non-anomalous conditions) is used to train unsupervised models, and the set of training data collected from device $e_i$ for $T$ timesteps can be defined as matrix $D_{e_i}$,
\begin{equation}
D_{e_i} = [d_{m,t}]_{M\times T}
\end{equation}

For each edge device $e_i$, the Model per Device (MPD) baseline approach takes $D_{e_i}$ as input and trains a model by minimizing the loss function $L$ to find the optimal model parameters $w^{mpd}_{e_i}$, as shown in equation \ref{equation:mpd}. Since a model is trained per edge device and microservices within a specific edge device are expected to have similar normal metric distributions (due to the QoS and resource requirement-aware placement of microservices) \cite{pallewataa2022qos, fu2021qos}, the accuracy of these models is expected to be high. However, because a large number of models need to be trained, this approach is not very efficient.

\begin{equation}
\label{equation:mpd}
w^{mpd}_{e_i} = arg min_w L(w;D_{e_i})
\end{equation}

On the other hand, the generic model (GM) baseline approach takes the data belonging to all edge devices, denoted as $D_E$, as input and trains a single model by minimizing the loss function $L$ to find the optimal model parameters $w^{gm}$, as shown in equation \ref{equation:gm}. A single generic model compromises accuracy in favor of training efficiency; a single model is unlikely to generalize well across the heterogeneous normal data distributions of all microservices in an edge environment, yet can be trained with fewer resources when compared to training multiple specialized models.

\begin{equation}
\label{equation:gm}
w^{gm} = arg min_w L(w;D_E)
\end{equation}

Since the above-explained baseline model training approaches are extremes in terms of model accuracy and training efficiency, it is important to reach a trade-off between accuracy and efficiency during the training process. This aim can be achieved by training a few models on groups of similar data, i.e., the heterogeneity of data used to train a single model should be minimal. To accomplish this, we use the concept of clustering to group edge devices with similar normal data distributions and then train the models within these clusters. 
Clustering is performed based on edge devices since the smallest unit of model deployment is the edge device level. Towards that, in section \ref{subsec:simcalc}, we introduce a similarity metric capable of clustering edge devices with similar normal data distributions, followed by sections \ref{subsec:icptl} and \ref{subsec:cm}, where we propose the two clustering-based training approaches, IPCTL and CM, respectively. 

\subsection{Similarity-based clustering}
\label{subsec:simcalc}

In this context, we assume that microservices are placed based on their QoS and resource requirements. As a result, microservices deployed at different hierarchical levels show a variance in metrics. This variance is more pronounced in some metrics (e.g., CPU and memory consumption metrics) than in others (e.g., error counts, disk read/write throughput). For instance, microservices situated closer to the cloud generally have higher values for CPU and memory consumption metrics and lower values for latency than those closer to the IoT layer. Hence, we identify a subset of $H$ such metrics, representative of the QoS and resource usage of microservices, suitable to perform device clustering from the $M$ metrics.

The first step towards cluster formation is calculating the similarity between each pair of devices, as proposed in Algorithm \ref{algo:SimilarityCalcAlgorithm}. The proposed algorithm takes as input normal datasets collected from all edge devices $D_E = \{D_{e_1},D_{e_2},..,D_{e_i},..,D_{e_N}\}$ and the list of $H$ representative metrics. It first extracts the $H$ representative metrics from each dataset $D_{e_i}$, resulting in $D_{e_i}^* = [d_{h,t}]_{H\times T}$. Next, each metric of $D_E^*$ is standardized using min-max normalization as shown in step \ref{sim_calc:standardize} of Algorithm \ref{algo:SimilarityCalcAlgorithm}. 
Subsequently, the algorithm generates a multivariate probability distribution $P_{e_i}$ composed of the probability distributions for each of the $H$ representative metrics, for each device $e_i$.

\label{sim_calc}
\begin{algorithm}[th]
	\caption{Similarity Graph Formation Algorithm} 
	\label{algo:SimilarityCalcAlgorithm}
	\begin{algorithmic}[1]
	    \STATE{\textit{Input} : Normal datasets collected from all edge devices $D_E = \{D_{e_1},D_{e_2},..,D_{e_i},..,D_{e_N}\}$}
     \STATE{\textit{Input} : List of $H$ representative metrics}
            \STATE{\textit{Output} : Similarity graph $SG$}
    	\STATE{Compose $D_E^* = \{D_{e_1}^*,D_{e_2}^*,..,D_{e_i}^*,..,D_{e_N}^*\}$ (where $D_{e_i}^* = [d_{h,t}]_{H\times T}$) by extracting the $H$ representative metrics}
            \STATE{\label{sim_calc:standardize} Standardize $D_E^*$ such that \\ $D_E^*[h,:] \gets \frac{D_E^*[h,:]}{max(D_E^*[h,:])-min(D_E^*[h,:])},\quad \forall h; 0 < h\le H$}
	    \STATE{Obtain $P_E = \{P_{e_1},P_{e_2},..,P_{e_i},..,P_{e_N}\}$ using $D_E^*$, where $P_{e_i}=[p_{e_{i,1}},p_{e_{i,2}},..,p_{e_{i,H}}]$ is the multivariate probability distribution for each edge device $e_i$}
	    \STATE Initialize similarity graph $SG=(\mathbb{V},\mathbb{E})$, where $\mathbb{V}=[e_i,\quad \forall i; 0<i \le N]$ and $\mathbb{E} = \emptyset$
            \FOR{each device $e_i ; 0<i\le N$}
            \FOR{each device $e_j ; i<j\le N$}
	    \STATE{$sim\_dist=\sqrt{\sum_{h=1}^{H} JS\_distance(p_{e_{i,h}},p_{e_{j,h}})^{2}}$}
            \STATE{$\mathbb{E} \leftarrow \mathbb{E} \bigcup \{(e_i,e_j,sim\_dist)\}$}
            \ENDFOR
            \ENDFOR
        \STATE{\textit{Return} : $SG$}
	\end{algorithmic}
\end{algorithm}

Once the multivariate probability distributions for all devices are available, the similarity distance $sim\_dist(e_i,e_j)$ between each pair of devices $e_i$ and $e_j$ is calculated by obtaining the L2-norm of the $JS\_distance$ between each representative metric of the probability distributions $P_{e_i}$ and $P_{e_j}$ corresponding to those devices. The equation corresponding to this step is shown in equation \ref{equation:sim_metr}. 
The $JS\_distance$ between the probability distributions $p_{e_{i,h}}$ and $p_{e_{j,h}}$ corresponding to a single metric $h$, is obtained by calculating the square root of their $JS\_divergence$, as shown in equation \ref{equation:jsd}. The $JS\_divergence$ between any two probability distributions, $\mathbb{P}$ and $\mathbb{Q}$, is the weighted sum of the forward and backward Kullback-Leibler (KL) divergence between those probability distributions (refer to equation \ref{equation:jsdiv_general}). The equation for calculating the KL divergence $KL(\mathbb{P} \parallel \mathbb{Q})$ between any two probability distributions, $\mathbb{P}$ and $\mathbb{Q}$, is shown in equation \ref{equation:kld_general} \cite{mlmastery2019kldiverg}.

\begin{equation}
\label{equation:sim_metr}
\begin{split}
\hspace*{-0.4in} sim\_dist(e_i,e_j) = \sqrt{\sum_{h=1}^{H} JS\_distance(p_{e_{i,h}},p_{e_{j,h}})^{2}}
\end{split}
\end{equation}

\begin{equation}
\label{equation:jsd}
\begin{split}
JS\_distance(p_{e_{i,h}},p_{e_{j,h}}) = \sqrt{JS\_divergence(p_{e_{i,h}}||p_{e_{j,h}})}
\end{split}
\end{equation}

\begin{equation}
\label{equation:jsdiv_general}
\begin{split}
\hspace*{-0.4in} JS\_divergence(\mathbb{P}||\mathbb{Q}) = \frac{1}{2}KL\left(\mathbb{P}||\frac{\mathbb{P}+\mathbb{Q}}{2}\right) \\+ \frac{1}{2}KL\left(\mathbb{Q}||\frac{\mathbb{Q}+\mathbb{P}}{2}\right)
\end{split}
\end{equation}

\begin{equation}
\label{equation:kld_general}
\begin{split}
KL(\mathbb{P} \parallel \mathbb{Q}) = \sum_{i} \mathbb{P}(i) \log \left( \frac{\mathbb{P}(i)}{\mathbb{Q}(i)} \right)
\end{split}
\end{equation}



The output of Algorithm \ref{algo:SimilarityCalcAlgorithm} is a complete graph, where a vertex represents an edge device $e_i$, and the weight of the edge between each pair of vertices $e_i$ and $e_j$ is the similarity distance $sim\_dist(e_i,e_j)$ between the corresponding edge devices. This graph is hereafter referred to as the similarity graph $SG$.

\begin{algorithm}[t]
	\caption{Similarity-based Device Clustering Algorithm} 
	\label{algo:KruskalClusteringAlgorithm}
	\begin{algorithmic}[1]
	    \STATE{\textit{Input} : Similarity graph $SG=(\mathbb{V},\mathbb{E})$ (output of Algorithm \ref{algo:SimilarityCalcAlgorithm})}
     \STATE{\textit{Input} : No.of clusters $K$}
            \STATE{\textit{Output} : Cluster to device mapping $cluster\_map$}
            \STATE{Initialize $cluster\_map = \{c_k:[e_k], \quad \forall k; 0<k \le N\}$}
            \WHILE{$length(cluster\_map)>K$}
                \STATE\label{krusk_clust:shortest_edge}{Find $e_{i,j}^{min} = (e_{i^*},e_{j^*},w_{{i^*},{j^*}})$ s.t.\\ $(e_{i^*},e_{j^*},w_{{i^*},{j^*}})=\underset{{(e_i,e_j,w_{i,j})\in \mathbb{E}}}{\operatorname{argmin}} w_{i,j}$}
                \STATE{$c_i \leftarrow cluster\_map^{-1}(e_i)$} // Look up the cluster to which the device $e_i$ belongs to
                \STATE{$c_j \leftarrow cluster\_map^{-1}(e_j)$}
                // Look up the cluster to which the device $e_j$ belongs to
                \STATE{$c_i \leftarrow c_i \bigcup c_j$}
                \STATE{$cluster\_map \leftarrow cluster\_map \setminus c_j$}
                \STATE\label{krusk_clust:last_step}{$\mathbb{E} \leftarrow \mathbb{E} \setminus e_{i,j}^{min}$}
            \ENDWHILE
            \STATE{\textit{Return} : $cluster\_map$}
	\end{algorithmic}
\end{algorithm}

The similarity graph $SG$ produced by Algorithm \ref{algo:SimilarityCalcAlgorithm} is then used to generate clusters of devices with similar normal data distributions, as depicted in Algorithm \ref{algo:KruskalClusteringAlgorithm}. Furthermore, the number of clusters $K$ is a hyperparameter 
and is required as input for this algorithm. In the first step, the $cluster\_map$, which stores the mapping from devices to clusters, is initialized such that each edge device $e_k$ is assigned to a unique cluster $c_k$. Therefore, initially, there are as many clusters as edge devices. Next, the algorithm identifies vertices corresponding to the shortest edge of the similarity graph $SG$, i.e., the edge devices $e_i$, $e_j$ with the highest similarity, and merges clusters containing those edge devices as depicted in steps \ref{krusk_clust:shortest_edge} to \ref{krusk_clust:last_step} of Algorithm \ref{algo:KruskalClusteringAlgorithm}. These steps are repeated until there are $K$ clusters in the $cluster\_map$. This algorithm, inspired by Kruskal's algorithm \cite{kleinberg2005algo_design}, ensures that clusters are created by grouping together devices corresponding to shorter edges. It returns the $cluster\_map$ with $K$ clusters as the output.


\begin{figure*}[th]
    \centering
    \subfigure[Generic model]{\includegraphics[width=.245\textwidth,
    height=3.5cm]{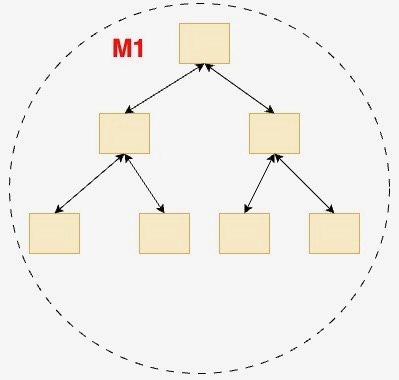}}
    \subfigure[Cluster-level model training]{\includegraphics[width=.245\textwidth, height=3.5cm]{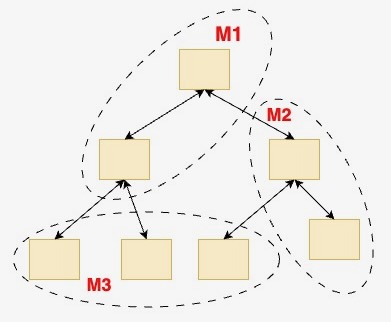}} 
    \subfigure[Intra-cluster parameter transfer learning-based model training]{\includegraphics[width=.245\textwidth, height=3.5cm]{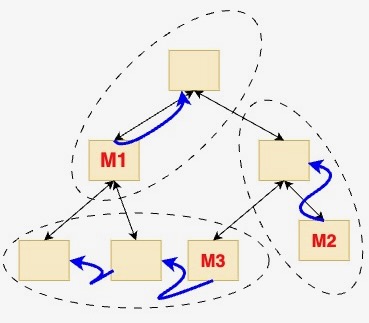}}
    \subfigure[Model per device]{\includegraphics[width=.245\textwidth, height=3.5cm]{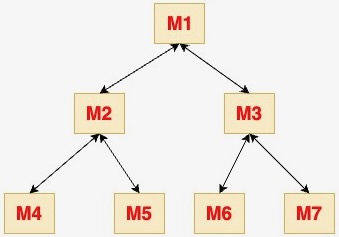}}
    \caption{Visual representation of the 2 clustering-based training approaches and the 2 baseline approaches}
    \label{visual_representation}
\end{figure*}

\subsection{Intra-cluster parameter transfer learning-based model training}
\label{subsec:icptl}
\begin{algorithm}[tb]
	\caption{Intra-cluster Parameter Transfer Learning-based Model Training Algorithm} 
	\label{algo:paramTLalgorithm}
	\begin{algorithmic}[1]
	    \STATE{\textit{Input} : Similarity graph $SG=(\mathbb{V},\mathbb{E})$ (output of Algorithm \ref{algo:SimilarityCalcAlgorithm})}
     \STATE{\textit{Input} : Cluster to device mapping $cluster\_map$ (output of Algorithm \ref{algo:KruskalClusteringAlgorithm})}
            \STATE{\textit{Output} : Models for all devices}
            \FOR{each cluster $c_k ; 0<k\le K$}
            \STATE{Initialize $source\_set = \emptyset$} \STATE{Initialize $target\_set = cluster\_map[c_k]$}
	    \STATE{Extract $SG_k = (\mathbb{V}_k,\mathbb{E}_k)$ where $\mathbb{V}_k=cluster\_map[c_k]$ and $\mathbb{E}_k=\{(e_i,e_j,w_{i,j})\in\mathbb{E}; e_i, e_j\in\mathbb{V}_k\}$}
    	\STATE{Find $e_{i,j}^{min} = (e_{i^*},e_{j^*},w_{{i^*},{j^*}})$ s.t.\\ $(e_{i^*},e_{j^*},w_{{i^*},{j^*}})=\underset{{(e_i,e_j,w_{i,j})\in \mathbb{E}_k}}{\operatorname{argmin}} w_{i,j}$}    
            \STATE{$source \leftarrow e_i$}
            \STATE{$target \leftarrow e_j$}
            \STATE\label{icptl_algo:mpd}{$w_{source} = \text{argmin}_w L(w;D_{source})$}
	    \STATE{$source\_set \leftarrow source\_set \bigcup source$}
            \STATE{$target\_set \leftarrow target\_set \setminus source$}
            \WHILE{$target\_set \neq \emptyset$}
                \STATE{\label{icptl_algo:icptl}$w_{target} = \text{argmin}_w L(w;D_{target},w_{source})$}
                \STATE{$source\_set \leftarrow source\_set \bigcup target$}
                \STATE{$target\_set \leftarrow target\_set \setminus target$}
                \STATE{$\mathbb{E}_k \leftarrow \mathbb{E}_k \setminus e_{i,j}^{min}$}
                \STATE{\label{icptl_algo:short_edge}Find $e_{i,j}^{min} = (e_{i^*},e_{j^*},w_{{i^*},{j^*}})$ s.t.\\ $(e_{i^*},e_{j^*},w_{{i^*},{j^*}})=\underset{{(e_i,e_j,w_{i,j})\in \mathbb{E}_k}}{\operatorname{argmin}} w_{i,j}$ and \\$e_i \in source\_set$}
                \STATE{$source \leftarrow e_i$}
                \STATE{$target \leftarrow e_j$}
            \ENDWHILE
            \ENDFOR
        \STATE{\textit{Return} : Models for all devices}
	\end{algorithmic}
\end{algorithm}

The aim of the intra-cluster parameter transfer learning-based model training (ICPTL) approach is to achieve a similar level of accuracy as the MPD approach but with fewer training cycles. In the MPD approach, each anomaly detection model needs to learn the normal data distribution from its respective edge device. However, the normal data distributions among edge devices within the same cluster only differ slightly. Therefore, in the ICPTL approach, instead of training the model for each edge device from scratch, we utilize the concept of parameter transfer learning, which involves transferring the weights learned by one model (the source model) and its hyperparameters to initiate the training of a model for another edge device (the target model) \cite{pan2010tl_survey}. For instance, if the source device is $e_i$ and the target device is $e_j$, the ICPTL approach takes dataset $D_{e_j}$ and model parameters of source model $w_{e_i}$ as inputs and trains a target model by minimizing the loss function $L$ to find the optimal model parameters $w^{icptl}_{e_j}$, as shown in equation \ref{equation:icptl}. This allows the target model to have an advantage at the beginning of its training, as the source model has already learned the lower-level features required by the target model, thus reaching a satisfactory accuracy within a few epochs.

\begin{equation}
\label{equation:icptl}
w^{icptl}_{e_j} = arg min_w L(w;D_{e_j},w_{e_i})
\end{equation}

Algorithm \ref{algo:paramTLalgorithm} comprises the steps of the ICPTL approach. It consumes the similarity graph $SG$, which is the output of Algorithm \ref{algo:SimilarityCalcAlgorithm} and the cluster to device mapping $cluster\_map$ which is the output of Algorithm \ref{algo:KruskalClusteringAlgorithm} as inputs. For a given cluster $c_k$, the $source\_set$, which stores the source devices with already trained models
and the $target\_set$, which stores the devices where the models are yet to be trained
are first initialized. Next, the subgraph $SG_k$, which consists of edge devices in $c_k$ as vertices, is extracted from the similarity graph $SG$. From $SG_k$, one of the vertices corresponding to the shortest edge, i.e., one of the edge devices $e_i$ from the pair exhibiting the highest similarity, is arbitrarily selected as the $source$ device. Next, a model is trained for that device by following the MPD approach as shown in step \ref{icptl_algo:mpd} of the algorithm. After training, the $source$ device $e_i$ is moved to the $source\_set$. Thereafter, the model for the remaining device $e_j$ (i.e., the $target$ device) is trained using parameter transfer learning as shown in equation \ref{equation:icptl}, and it also corresponds to step \ref{icptl_algo:icptl} of the algorithm. After this step, the $target$ device $e_j$ is also moved to the $source\_set$. Subsequently, as shown in step \ref{icptl_algo:short_edge}, the algorithm identifies the next shortest edge from the remaining edges, such that the edge device corresponding to one of its vertices belongs to the $source\_set$. Next, by utilizing the model for that edge device as the source model, the model for the edge device corresponding to the other vertex is trained using parameter transfer learning. These steps are repeated until the $target\_set$ becomes empty. At the end of the algorithm, a trained model will be available for all edge devices within $c_k$. Moreover, ICPTL-based model training can be performed simultaneously across clusters.

Figure \ref{visual_representation}(c) visually represents the ICPTL approach. It essentially identifies a Minimum Spanning Tree (MST) of the cluster as the sequence to perform parameter transfer learning among the edge devices. In the identified sequence, a model is traditionally trained only for the device at the starting point of the sequence, while the other devices in the sequence receive the model from the nearest neighbor device and train a few epochs until they reach satisfactory accuracy. While this approach entails training a model for each edge device, it requires fewer epochs during training compared to the MPD approach, thus resulting in reduced training resource requirements and training time.

\subsection{Cluster-level model training} 
\label{subsec:cm}
Through the cluster-level model training (CM) approach, we aim to improve training efficiency further by reducing the number of models. To that end, we propose to train a model per cluster using normal data collected from all edge devices in the cluster. The CM approach takes the data belonging to all edge devices in cluster $c_k$, denoted as $D_k$, as input and trains a model for that cluster by minimizing the loss function $L$ to find the optimal model parameters $w^{cm}_k$ as shown in equation \ref{equation:cm}. The CM approach is visually represented in Figure \ref{visual_representation}(b).

\begin{equation}
\label{equation:cm}
w^{cm}_k = arg min_w L(w;D_k)
\end{equation} 

Since similarity-based clustering ensures that devices having similar normal data distributions fall into one cluster, the model trained for each cluster can effectively generalize across all the edge devices in that cluster. This cluster-specific approach is expected to be more accurate than the generic model (GM) approach. 

Since the number of clusters is always less than or equal to the number of edge devices, this means that fewer models need to be trained when using the CM approach compared to the MPD approach. This results in reduced training time and resource requirements. Additionally, it is easier to manage a few models per cluster than a relatively large number of models per edge device. As a result, the CM approach is expected to be more efficient than the MPD approach.

\begin{figure}[th]
\centering
\includegraphics[width=3.5in]{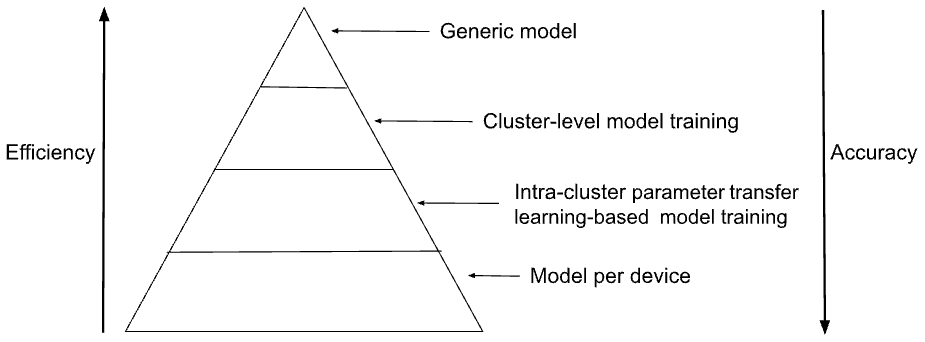}
\caption{Positioning of the two proposed clustering-based training approaches}
\label{proposed_approaches}
\end{figure}

Out of the two proposed approaches, CM is designed to provide more efficiency (in terms of training time, training resource requirements, number of epochs, and model management complexity) compared to ICPTL, while ICPTL is designed to provide more accuracy than CM. Figure \ref{proposed_approaches} depicts the positioning of the two proposed clustering-based training approaches amidst the two baseline model training approaches in terms of accuracy and efficiency.

\section{Performance Evaluation}
\label{sec:eval}

In this section, we discuss the evaluation results of the proposed clustering-based training approaches. First, we explain the details of the experimental setup used for evaluation. Following that, we compare four common unsupervised multivariate time-series anomaly detection techniques to identify the most suitable algorithm for the rest of our evaluations
. Finally, we discuss the results of evaluating the two proposed clustering-based training approaches using the algorithm identified from the previous step against the baseline approaches derived from the state of the art.

\subsection{Experimental Setup}

\begin{figure}[!t]
    \centering
    \subfigure{\includegraphics[width=\columnwidth]{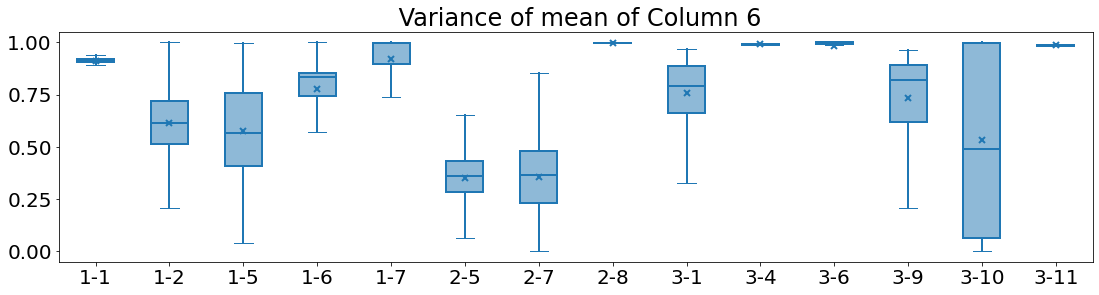}}
    \subfigure{\includegraphics[width=\columnwidth]{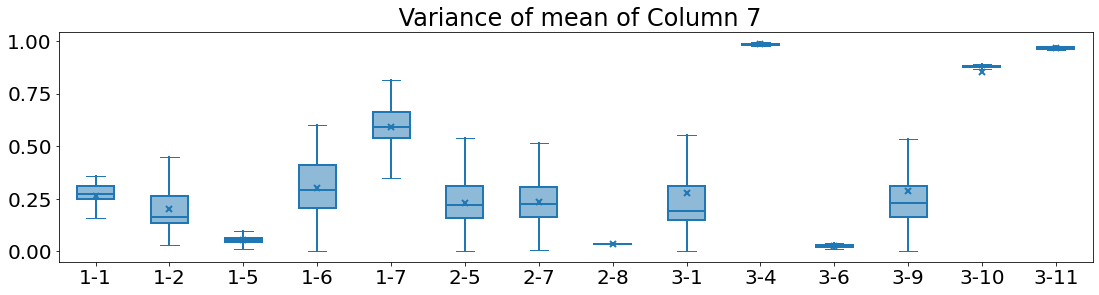}} 
    \subfigure{\includegraphics[width=\columnwidth]{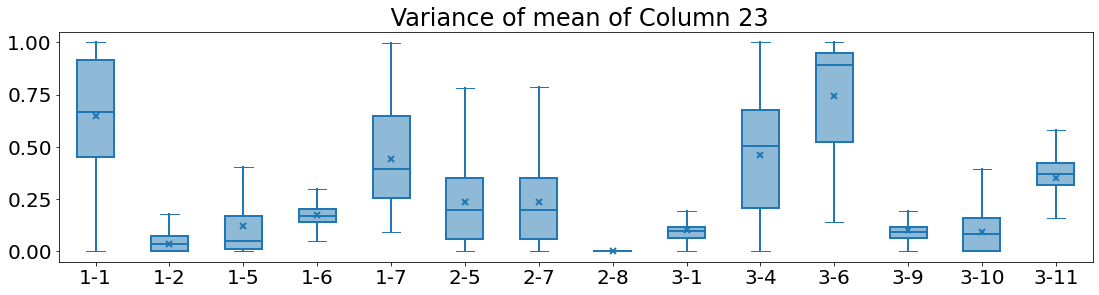}}
    \subfigure{\includegraphics[width=\columnwidth]{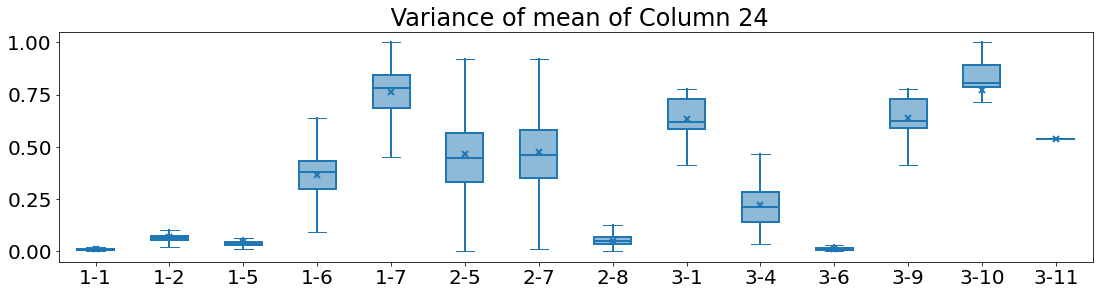}}
    \caption{Variance of mean of metrics across SMD devices}
    \label{fig:smdplot}
    \vspace{-15pt}
\end{figure}

In order to conduct a comprehensive and reproducible evaluation of the proposed approaches, we utilized a public performance anomaly dataset called the “Server Machine Dataset” (SMD), which is widely used in the anomaly detection literature \cite{su2019robust, audibert2020usad, audibert2022dodeep}. The SMD contains both normal and anomalous performance data from 28 devices across 38 metrics. Following the suggestion of Li et al. \cite{li2021intermetric}, we focused on 14 devices without a concept drift between training (normal performance) data and testing (anomalous performance) data for our evaluation. The SMD dataset has also been utilized in assessing research on edge computing environments \cite{tuli2023aiaugmented, tuli2021generative}, as the data from these devices exhibit heterogeneous normal data distributions akin to those in edge computing environments. This is apparent from the significant variance in mean values of certain selected metrics across the majority of devices, as shown in Figure \ref{fig:smdplot}.

In order to further verify the results obtained 
on SMD, we also collected performance anomaly data from a set of microservices with heterogeneous QoS and resource requirements deployed in an emulated edge computing environment. By extending the iContinuum \cite{negin2024icontinuum} emulation tool with anomalous data collection capabilities, we emulated 10 heterogeneous edge devices using 10 virtual machines (VMs) with different resource capacities (2vCPU/8GB, 4vCPU/16GB, 8vCPU/32GB) on Melbourne Research Cloud\footnote{https://dashboard.cloud.unimelb.edu.au/} and emulated the network (by following the communication link parameters used by Pallewatta et al.\cite{pallewataa2022qos} in their experiments) to represent the heterogeneity of network bandwidth between the edge devices. We deployed microservices with heterogeneous resource requirements representing a diverse set of IoT applications (from lightweight sensor data processing to heavy applications like camera face detection/recognition) on the emulated edge infrastructure by following the placement algorithm proposed by Pallewatta et al.\cite{pallewatta2023placement}. Using Pixie\footnote{https://docs.px.dev/}, which is a lightweight monitoring tool suitable for edge monitoring, we collected both normal performance data as well as anomalous performance data under a diverse set of anomalies, such as resource saturation, service failures, network saturation, etc., at a granularity of 10 seconds. Normal workloads were generated using Jmeter\footnote{https://jmeter.apache.org/} while anomalies were injected using Chaos Mesh\footnote{https://chaos-mesh.org/}. Both normal and anomalous performance data are available for 10 metrics, covering both system and application metrics.  

Experiments on both datasets, including hyperparameter tuning, were conducted on the Spartan\footnote{https://dashboard.hpc.unimelb.edu.au/} HPC cluster. More details on training, including the hyperparameters selected for training, are explained in sections \ref{subsec:algo_eval}, \ref{subsec:smd_cluster_eval}, and \ref{subsec:emu_cluster_eval}.

\subsection{Evaluating performance anomaly detection algorithms against the SMD dataset}
\label{subsec:algo_eval}

Prior to evaluating the proposed clustering-based training approaches using the SMD dataset, in this section, we compare four common unsupervised multivariate time-series anomaly detection techniques to identify the most suitable algorithm for the rest of our evaluations in terms of accuracy and efficiency during inference.

As mentioned in section \ref{subsec:perf_ad_edge}, while research on AI-based performance anomaly detection in cloud environments is well-established \cite{kardani2019performance, audibert2020usad, naikade2020automated, audibert2022dodeep}, studies on performance anomaly detection for edge environments have also based their works on cloud anomaly detection approaches \cite{becker2020towards, Soualhia2019infrastructure, Skaperas2024pragmatical}. According to Audibert et al., existing anomaly detection algorithms can be broadly classified into three categories: (1) conventional methods, (2) machine learning (ML)-based methods, and (3) deep neural network (DNN)-based methods \cite{audibert2022dodeep}. Therefore, we select the following four representative performance anomaly detection algorithms from each of these categories, and evaluate them to identify the most suitable algorithm for the rest of our evaluations.

\begin{enumerate}
\item AutoEncoder (AE) - DNN-based methods 
\item Isolation Forest (IF) - ML-based methods
\item LSTM-AutoEncoder (LSTM-AE) - DNN-based methods
\item Vector AutoRegression (VAR) - conventional methods
\end{enumerate}

Since the VAR algorithm is one of the most commonly used conventional methods in multivariate time series anomaly
detection \cite{audibert2022dodeep}, we chose VAR algorithm and implemented it using the Statsmodels\footnote{https://www.statsmodels.org/} library. IF is a popular ML-based algorithm considered in multiple anomaly detection works \cite{kardani2019performance, audibert2020usad, su2019robust, tuli2021generative}. The original IF approach in the Scikit-learn library for Isolation Forest\footnote{https://scikit-learn.org} only takes the spatial dependencies among data points into consideration when performing anomaly detection. However, we modified it following Kardani et al. \cite{kardani2019performance} who have incorporated the k-point moving average into the feature space of each sample. It captures the temporal dependencies of the performance data by obtaining the average of a window of k-previous samples. We selected two algorithms representing DNN-based methods: AE and LSTM-AE. AE, as well as approaches that have extended the AE concept \cite{audibert2020usad, Zong2018DAGMM, tuli2021generative}, are a class of popular DNN-based methods. In addition to AE, we also considered LSTM-AE, since LSTM layers are capable of capturing time-series dependencies better \cite{su2019robust, naikade2020automated, becker2020towards, audibert2022dodeep}. The AE and LSTM-AE ML models were implemented using the Pytorch\footnote{https://pytorch.org/} library. We employed the Tree-structured Parzen Estimator \cite{bergstra2011tpe}, which is a Bayesian optimization technique, to tune the hyperparameters listed in Table \ref{table:hyperparam_table} corresponding to each algorithm.
\begingroup
\setlength{\tabcolsep}{2pt} 
\renewcommand{\arraystretch}{1} 
\begin{table}[t]
\caption{Hyperparameters tuned for each anomaly detection algorithm\label{table:hyperparam_table}}
\small
\begin{tabular*}{\columnwidth}{p{0.2\columnwidth}|p{0.8\columnwidth}}
\hline
\textbf{Machine Learning Model} & \textbf{Tuned Hyperparameters} \\
\hline
AE & num\_layers, window\_size, hidden\_size, batch\_size, learning\_rate\\
IF & n\_estimators, max\_features, max\_samples, k (for the k-point moving average)\\
LSTM-AE & num\_layers, window\_size, batch\_size, learning\_rate\\
\hline
\end{tabular*}
\end{table}
\endgroup
\begingroup
\setlength{\tabcolsep}{2pt}
\renewcommand{\arraystretch}{1} 
\begin{table}[t]
\caption{Evaluation results (F1-score, AUC) of anomaly detection algorithms
\label{table:cloud_approach_eval}}
\small
\begin{tabular*}{\columnwidth}{p{0.2\columnwidth}|p{0.2\columnwidth}|p{0.2\columnwidth}|p{0.2\columnwidth}| p{0.2\columnwidth}}
\hline
 & \textbf{AE} & \textbf{IF} & \textbf{LSTM-AE} & \textbf{VAR}\\\cline{2-3}
\hline
AUC & \textbf{0.89} & 0.84 & 0.86 & 0.74\\
F1-score & \textbf{0.79} & 0.69 & 0.73 & 0.50\\
\hline
\end{tabular*}
\end{table}
\endgroup

\begin{figure}[th]
\centering 
\includegraphics[width=0.5\textwidth]{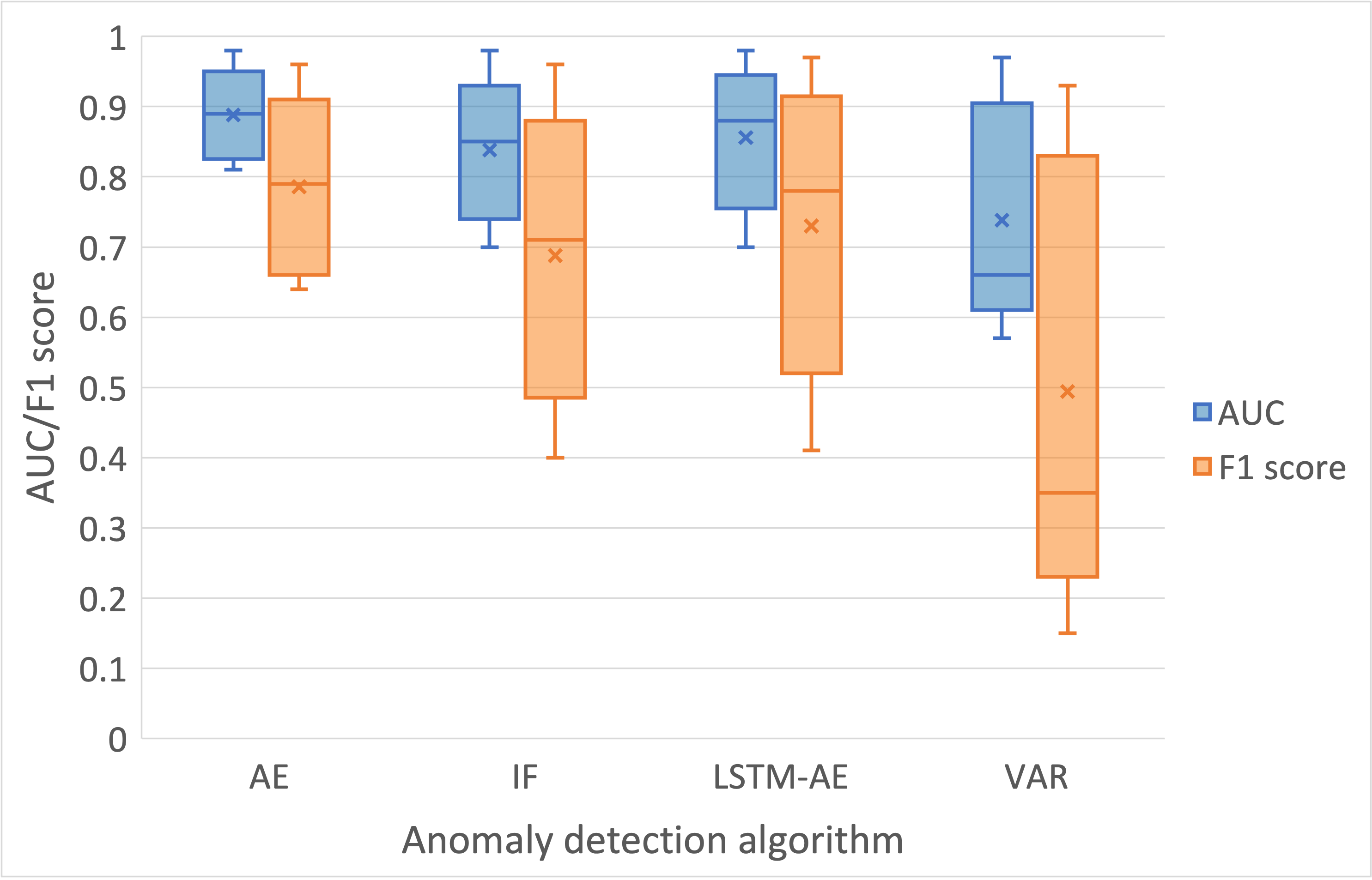}
	\caption{Distribution of AUC and F1 score of anomaly detection algorithms across devices in SMD}
	\label{fig:boxplot_for_algos}
\end{figure}

We use F1-score and AUC to evaluate the accuracy of performance anomaly detection algorithms similar to other works in the domain \cite{kardani2019performance, matsuo2021performance}. F1-score is the harmonic mean between precision and recall, while AUC refers to the area under the receiver operating characteristic curve. 

We evaluate the four anomaly detection algorithms on a random subset of devices of the SMD dataset by following the MPD approach. Table \ref{table:cloud_approach_eval} shows the evaluation results in terms of average AUC and F1-scores. Figure \ref{fig:boxplot_for_algos} shows the distribution of AUC and F1-scores across devices for each anomaly detection algorithm. AE achieves the highest mean AUC (0.89) and F1-score (0.79) out of all algorithms. As shown in Figure \ref{fig:boxplot_for_algos}, AE also has the lowest variance. 

During our evaluation, we consider both the accuracy and efficiency of the four algorithms. The VAR algorithm does not require a training phase, but it has a high inference time. This results in high latency when detecting anomalies. It also has the least mean and highest variance for its AUC and F1-scores among the compared algorithms. Therefore VAR is the least suitable anomaly detection algorithm in terms of accuracy and efficiency. IF is a low overhead algorithm with low linear-time complexity and small memory requirements. However, IF-based approaches require a certain percentage of anomaly data for training to achieve their best possible accuracy \cite{kardani2019performance}. This is also evident through the low mean and high variance of AUC and F1-scores obtained in this evaluation, where the IF algorithm was trained only on normal data. Out of the remaining two algorithms, AE has the best mean and least variance for its AUC and F1-scores. LSTM-AE also has a nearly equal mean but higher variance on its AUC and F1-scores than AE. LSTM-AE does not outperform the AE algorithm, although it is designed to capture time-series dependencies. Out of these two DNN-based approaches, when considering efficiency, the AE model is less complex, and its inference can be parallelized, unlike an LSTM-AE model. Therefore, we conclude that AE is the most suitable performance anomaly detection algorithm for the rest of our evaluations. AE models used in the rest of the evaluations are trained by minimizing the Mean Squared Error (MSE) loss between the original and reconstructed windows using the Adam optimizer. However, the proposed clustering-based training approaches are independent of the ML approach. Hence, a reader can use any other independent model to evaluate the proposed training approaches.

\subsection{Evaluating clustering-based training approaches using the SMD dataset}
\label{subsec:smd_cluster_eval}

\begin{figure}[th]
	\centering 
	\includegraphics[width = \columnwidth]{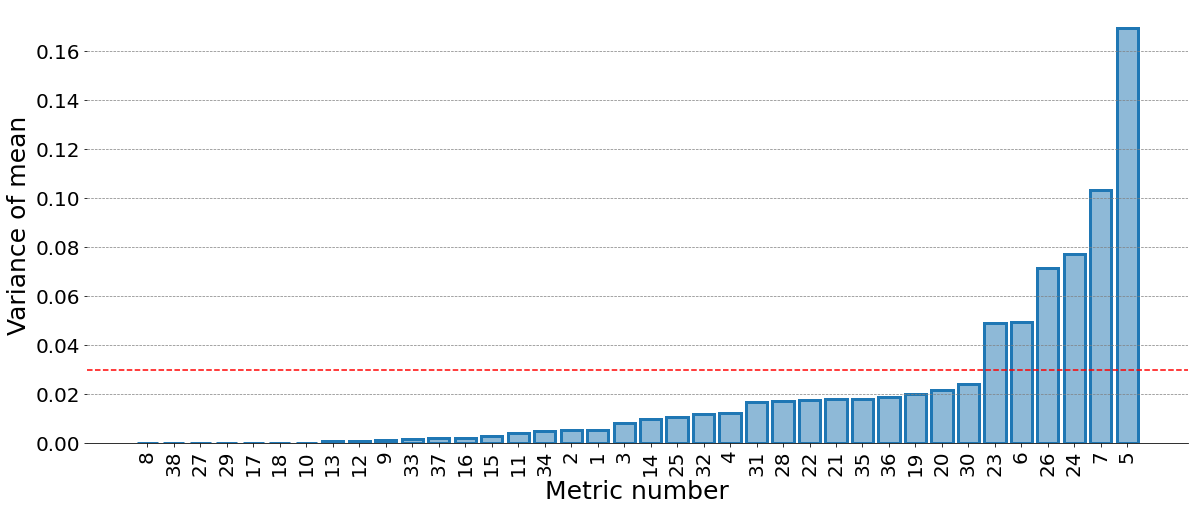}
	\caption{Variance of mean of each metric across SMD devices}
	\label{fig:variance_of_mean}
\end{figure}

In this section, we evaluate the proposed clustering-based training approaches on the SMD dataset. As the first step, we assign edge devices to clusters. However, since metrics are undefined in this dataset, we can not use domain knowledge to select the subset of metrics suitable for performing device clustering. Therefore, we use the variance of mean to identify significantly varying metrics across devices to identify the subset of metrics. The variance of mean represents how much the mean of each metric varies across devices. Figure \ref{fig:variance_of_mean} is a bar chart that represents the variance of mean of each metric across devices. As can be seen, columns 23, 6, 24, 26, 7, and 5 have a high variance of mean. However, column 5 has all-zero occurrences in 9 out of 14 devices. Hence, we leave it out of consideration for clustering.
\begin{figure}[t]
	\centering 
	\includegraphics[width=\columnwidth]{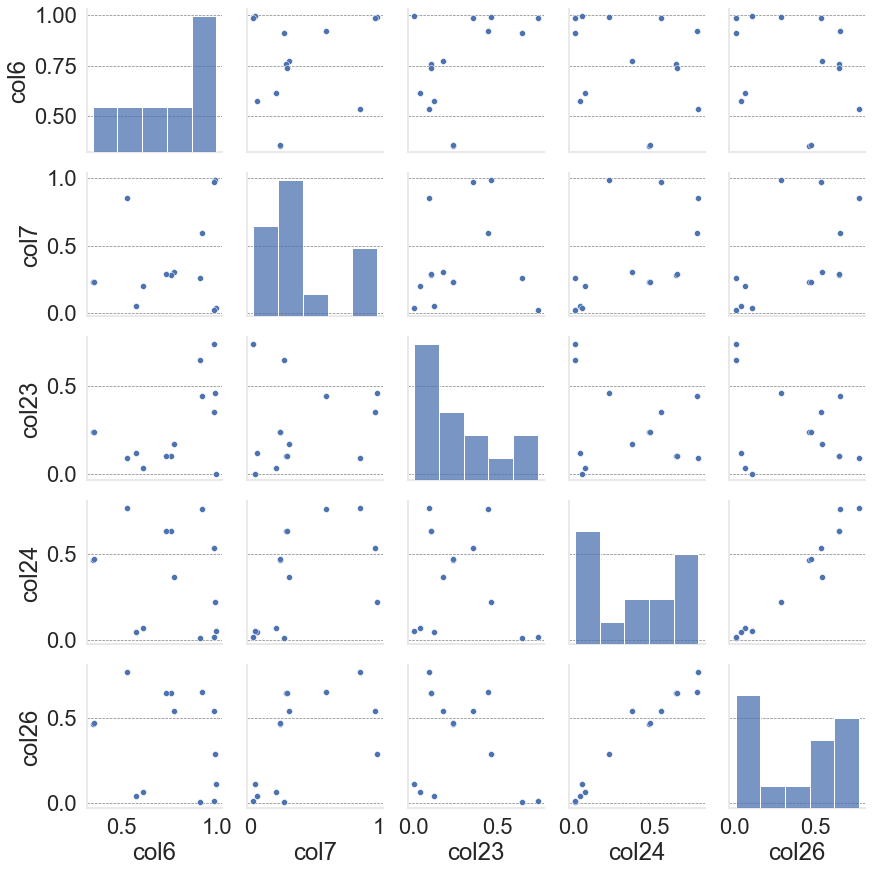}
	\caption{Collinearity between shortlisted metrics across SMD devices}
	\label{fig:pairplots}
 \vspace{-15pt}
\end{figure}

Next, we obtain the collinearity between those metrics. Figure \ref{fig:pairplots} contains the pairplot, which shows the collinearity between metrics shortlisted from the previous step. Since 24 and 26 are linearly correlated, one of those (26) was dropped from the set of metrics for clustering.

Then, we clustered the 14 devices based on the final set of metrics using Algorithms \ref{algo:SimilarityCalcAlgorithm} and \ref{algo:KruskalClusteringAlgorithm}. All the algorithms corresponding to similarity-based clustering and clustering-based training approaches were implemented using Python. The same AE model implementation mentioned in section \ref{subsec:algo_eval} was used for evaluations in this section as well. We employed the Tree-structured Parzen Estimator \cite{bergstra2011tpe} Bayesian optimization technique, to optimize the hyperparameters of the AE model listed in Table \ref{table:hyperparam_table} as well as to determine the optimal number of clusters ($K$) for Algorithm \ref{algo:KruskalClusteringAlgorithm}. Subsequent to hyperparameter tuning, the optimal number of clusters, $K$ was identified to be five.

Next, we present the evaluation results for the two proposed clustering-based training approaches.

\begingroup
\setlength{\tabcolsep}{2pt}
\renewcommand{\arraystretch}{1} 
\begin{table}[t]
\caption{Evaluation results of training approaches against SMD
\label{table:training_approach_eval_smd}}
\small
\begin{tabular*}{\columnwidth}{p{0.27\columnwidth}|p{0.2\columnwidth}|p{0.17\columnwidth}|p{0.18\columnwidth}| p{0.15\columnwidth}}
\hline
 & \textbf{MPD} & \textbf{ICPTL} & \textbf{CM} & \textbf{GM}\\\cline{2-3}
\hline
AUC (mean) & 0.88 & 0.85 & 0.81 & 0.66\\
F1-score (mean) & 0.77 & 0.7 & 0.65 & 0.35\\
Total training time & 19min 40secs & 8min 2secs & 6min 8secs & 47secs\\
\hline
\end{tabular*}
\end{table}
\endgroup

\begin{figure}[th]
\centering 
\includegraphics[width=\columnwidth]{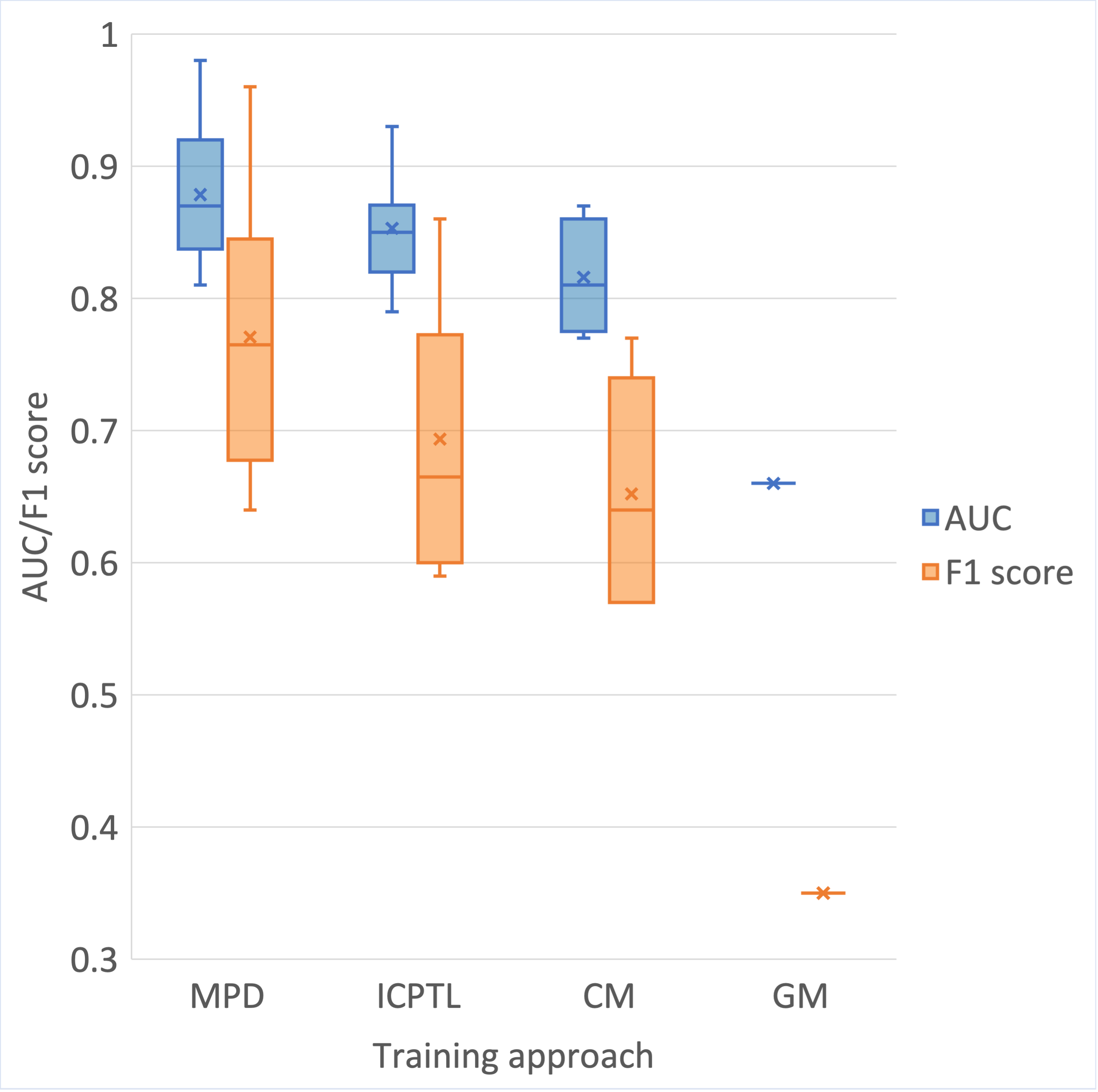}
	\caption{Distribution of AUC and F1 score of training approaches against SMD 
}
	\label{fig:boxplot_for_training}
\end{figure}

Table \ref{table:training_approach_eval_smd} contains the accuracy (in terms of average AUC and F1-scores) and training time results of the two proposed clustering-based model training approaches and the two baseline approaches. Figure \ref{fig:boxplot_for_training} shows the distribution of AUC and F1-scores across these training approaches. As predicted, the MPD approach yielded the highest average AUC (0.88) and F1-score (0.77), but it required the longest total training time (19min 40secs). On the other hand, the GM approach had the lowest average AUC (0.66) and F1-score (0.35) with the shortest total training time (47secs). As expected, the ICPTL approach had the second-highest mean AUC (0.85) and F1-score (0.7) while consuming only 40\% of the total training time required by the MPD approach (8min 2secs). The CM approach has further increased training efficiency by consuming 23\% less training time than the ICPTL approach (6min 8secs) while reducing the total number of models from 14 to 5. Furthermore, the CM approach is 16\% and 30\% more accurate in AUC (0.81) and F1-score (0.65), respectively, than the GM approach.

\begin{figure*}[!t]
    \centering
    \subfigure[1-5 $\Rightarrow$ 1-2 (cluster 1)]{\includegraphics[width=.245\textwidth]{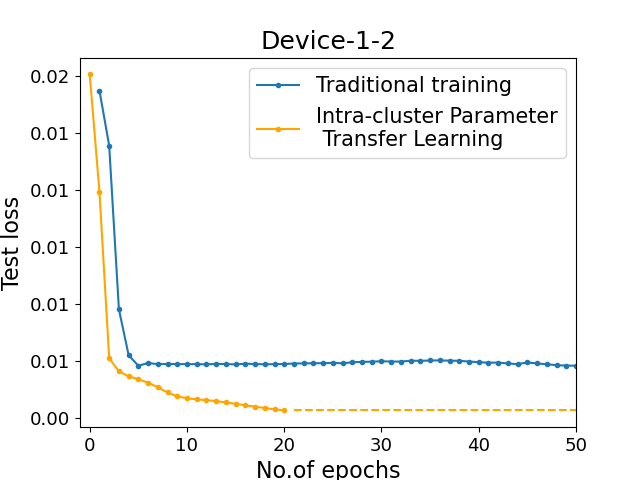}}
    \subfigure[3-9 $\Rightarrow$ 3-1 (cluster 2)]{\includegraphics[width=.245\textwidth]{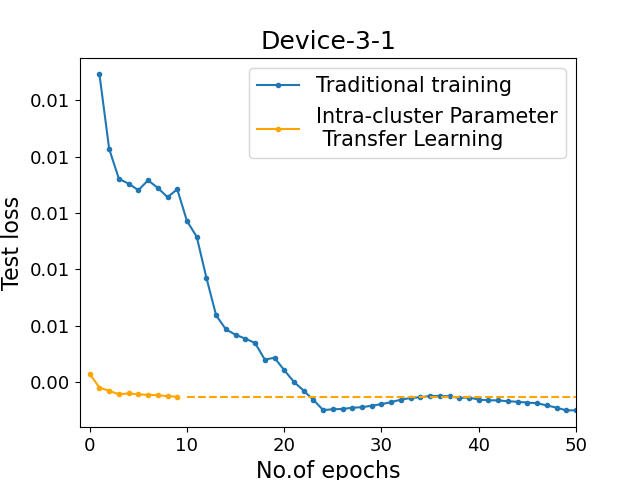}} 
    \subfigure[3-9 $\Rightarrow$ 1-7 (cluster 2)]{\includegraphics[width=.245\textwidth]{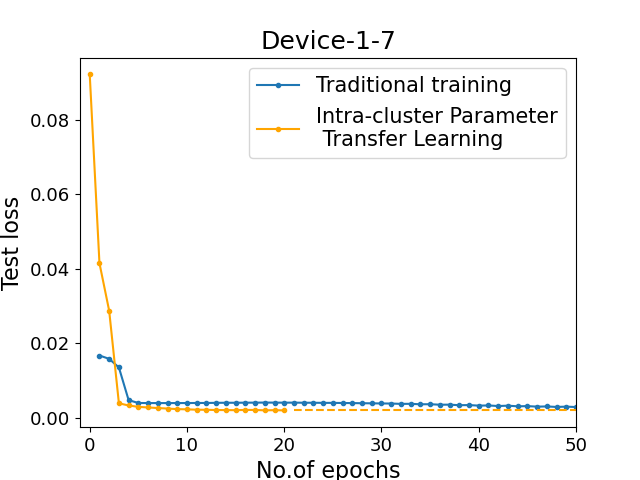}}
    \subfigure[2-5 $\Rightarrow$ 2-7 (cluster 0)]{\includegraphics[width=.245\textwidth]{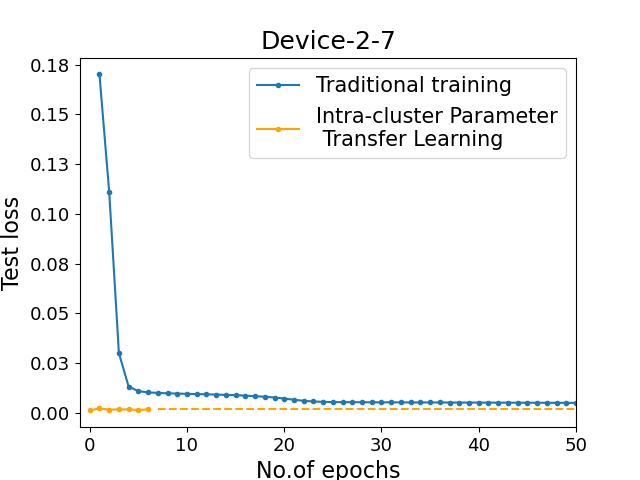}}
    \caption{Comparison between test loss of the ICPTL approach and the MPD approach for selected devices}
    \label{fig:cluster_for_testloss}
    \vspace{-15pt}
\end{figure*}

Next, we further explore how the ICPTL approach aims to achieve an accuracy similar to that of the MPD approach with fewer training cycles. In order to achieve this target, ICPTL should be able to converge to a lower test loss (similar to that of MPD) faster. Towards this, for a few selected devices, we compare the test loss for each epoch (for a duration of 50 epochs) in the ICPTL approach against that of the MPD approach. For instance, Fig.\ref{fig:cluster_for_testloss}a depicts the scenario where the model for "Device-1-2" was traditionally trained using the MPD approach as well as trained from the model for "Device-1-5" using the ICPTL approach. In some scenarios (e.g., Fig. \ref{fig:cluster_for_testloss}b, Fig. \ref{fig:cluster_for_testloss}d), the test set loss obtained with parameter transfer learned model (at its 0th epoch, i.e., without performing any retraining) is already low. Even in some cases where the test set loss of the parameter transfer learning approach is high (e.g., Fig. \ref{fig:cluster_for_testloss}a, Fig. \ref{fig:cluster_for_testloss}c), it converges faster within the first few epochs itself. Regardless of the initial test loss value, the ICPTL approach can converge to a test loss similar to that of the MPD approach faster, thus reaching a satisfactory accuracy within a few epochs, i.e., with no/minimal retraining. Although we have provided results for a few selected scenarios, we could observe the same behavior in other devices as well.

\begin{figure}[th]
\centering 
\includegraphics[width=\columnwidth]{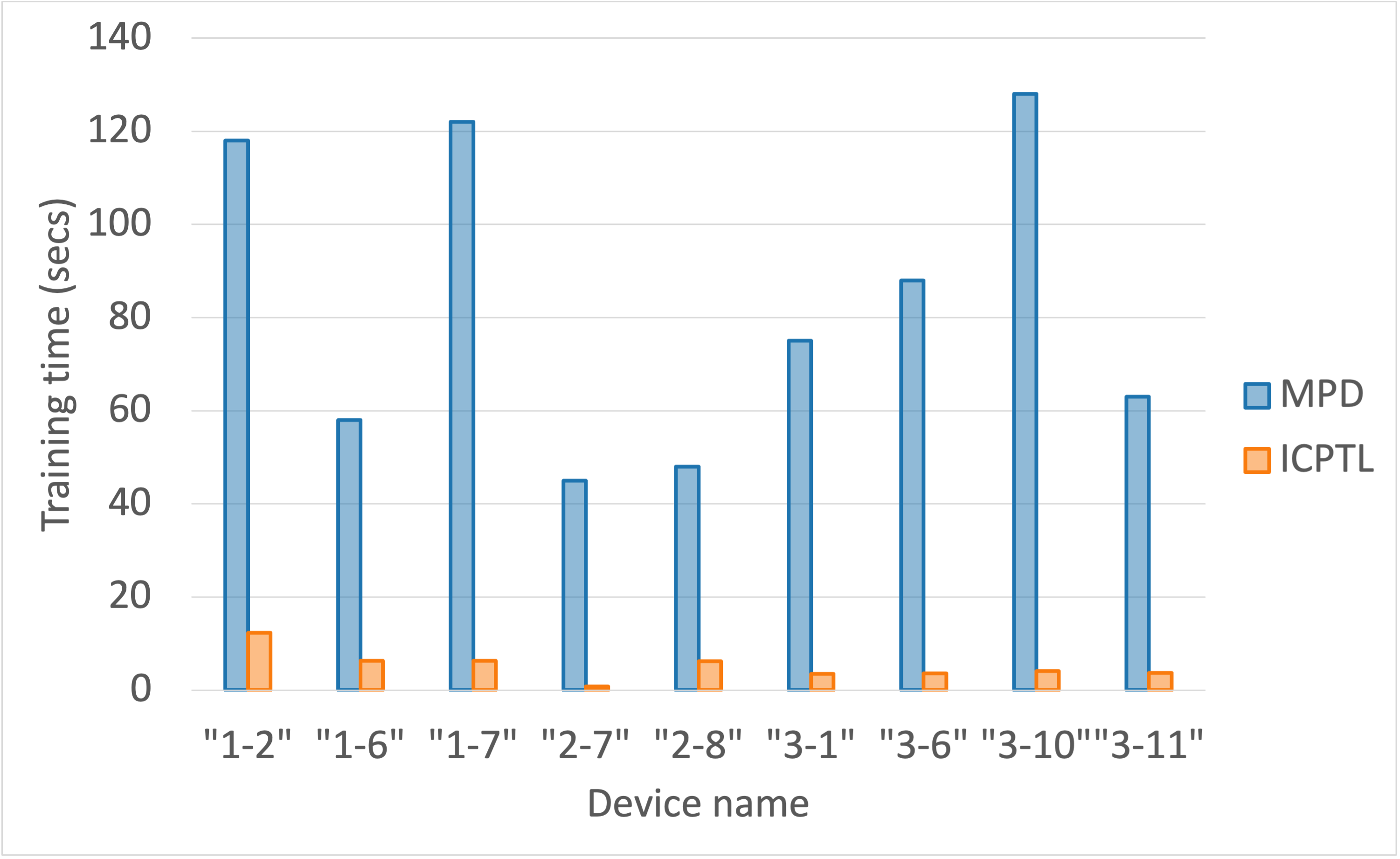}
	\caption{Training time comparison between ICPTL and MPD approaches against SMD}
	\label{fig:train_time_comp}
\end{figure}

For each scenario, the maximum number of epochs to perform parameter transfer learning is set to 20. We deployed an algorithmic stopping approach that stops training when validation loss does not change significantly for five epochs. As a result, in 4 out of 9 scenarios, parameter transfer learning stops before 20 epochs. Fig. \ref{fig:train_time_comp} presents a comparison between the training time results of the MPD approach and the ICPTL approach. In terms of training time efficiency, all nine parameter transfer learning scenarios were completed within a maximum of 12.33 seconds, whereas the corresponding traditional model training took a minimum of 45 seconds. While training the models for the nine devices using the MPD approach took a total of 12 minutes and 25 seconds, all nine parameter transfer learning scenarios completed training, taking only a total of 47 seconds, which is only 6\% of the training time required by the MPD approach. 

Based on our findings, the ICPTL approach achieves a comparable level of accuracy to the MPD approach but requires fewer training cycles. The CM approach demonstrates higher efficiency than the ICPTL approach in terms of total training time and model management complexity while still maintaining a higher accuracy than the GM approach. These results are further validated in the next section by conducting a small-scale experiment using data collected from the emulated edge computing environment.

\subsection{Evaluating clustering-based training approaches using the emulated dataset}
\label{subsec:emu_cluster_eval}

In this section, we further evaluate the proposed clustering-based training approaches using the dataset collected from the emulated edge computing environment. First, we clustered the 10 edge devices based on the collected system and application metrics using Algorithms \ref{algo:SimilarityCalcAlgorithm} and \ref{algo:KruskalClusteringAlgorithm}. We identified the optimal number of clusters, $K$ to be three through hyperparameter tuning.

\begingroup
\setlength{\tabcolsep}{2pt}
\renewcommand{\arraystretch}{1} 
\begin{table}[t]
\caption{Evaluation results of training approaches against the collected dataset
\label{table:training_approach_eval_our_dataset}}
\small
\begin{tabular*}{\columnwidth}{p{0.27\columnwidth}|p{0.2\columnwidth}|p{0.17\columnwidth}|p{0.16\columnwidth}| p{0.15\columnwidth}}
\hline
 & \textbf{MPD} & \textbf{ICPTL} & \textbf{CM} & \textbf{GM}\\\cline{2-3}
\hline
AUC (mean) & 0.88 & 0.88 & 0.86 & 0.73\\
F1-score (mean) & 0.95 & 0.96 & 0.95 & 0.9\\
Total training time & 13.98secs & 4.88secs & 3.8secs & 1.14secs\\
\hline
\end{tabular*}
\end{table}
\endgroup

\begin{figure}[th]
\centering 
\includegraphics[width=\columnwidth]{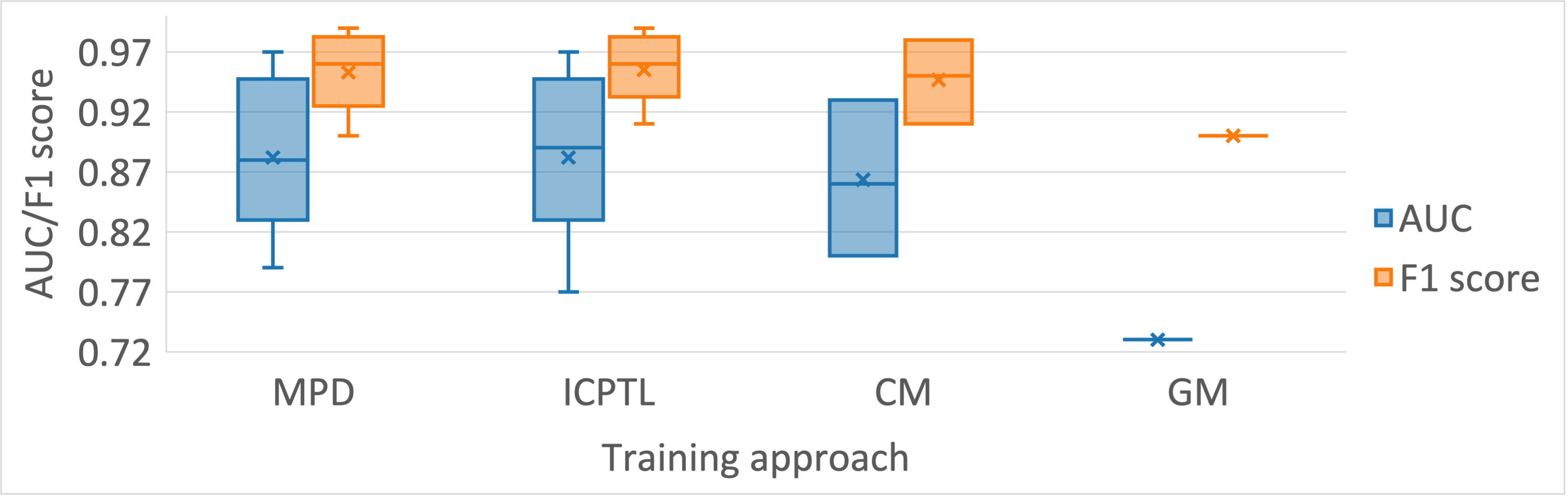}
	\caption{Distribution of AUC and F1 score of training approaches against the collected dataset 
}
	\label{fig:boxplot_for_training_our_dataset}
\end{figure}

Table \ref{table:training_approach_eval_our_dataset} contains the accuracy (in terms of average AUC and F1-scores) and training time results of the two proposed clustering-based model training approaches and the two baseline approaches. Figure \ref{fig:boxplot_for_training_our_dataset} shows the distribution of AUC and F1-scores across these training approaches. During this evaluation, it can be observed that the ICPTL approach had reached the accuracy of the MPD approach (which is also the highest accuracy: average AUC = 0.88 and average F1-score = 0.96) while consuming only 34\% of the total training time required by the MPD approach (4.88secs). The CM approach has further increased training efficiency by consuming 22\% less training time than the ICPTL approach (4.88secs) while reducing the total number of models from 10 to 3. Furthermore, the CM approach is 13\% and 5\% more accurate in AUC (0.86) and F1-score (0.95), respectively, than the GM approach. 

Thus, based on our experiment, we can assert that the ICPTL approach achieves the same accuracy as the MPD approach in a shorter training time. Furthermore, the CM approach is capable of reaching higher efficiency than the ICPTL approach while maintaining a greater accuracy than the GM approach.

\subsection{Theoretical comparison of clustering-based training approaches}

\begingroup
\setlength{\tabcolsep}{2pt}
\renewcommand{\arraystretch}{1} 
\begin{table}[t]
\caption{Theoretical comparison of efficiency metrics across all approaches ($1 \le K \le N$, $l<<L$)
\label{table:theoretical_comp}}
\small
\begin{tabular*}{\columnwidth}{p{0.16\columnwidth}|p{0.2\columnwidth}|p{0.5\columnwidth}}
\hline
\textbf{Approach} & \textbf{Model management complexity} & \textbf{Training resource requirements/ Training time}\\\cline{2-3}
\hline
GM & 1 & $\propto L$\\
CM & K & $\propto K*L$\\
ICPTL & N & $\propto K*L + (N-K)*l$\\
MPD & N & $\propto N*L$\\
\hline
\end{tabular*}
\end{table}
\endgroup

Apart from the empirical evaluation of training time in the previous sections, in Table \ref{table:theoretical_comp}, we present a theoretical comparison of other efficiency metrics, including training resource requirements and model management complexity, in addition to training time, across all approaches. Assuming that the dataset size provided for all the approaches is constant, we can discuss the training time and training resource requirements using the total number of epochs required during training. Considering that the number of devices is $N$ and the number of clusters is $K$ ($K \le N$), then MPD and CM approaches need to train $N$ and $K$ number of models, respectively. If the number of epochs is $L$, then the training time and training resource requirements to train all models in the MPD approach are proportional to $N * L$, and the training time and training resource requirements to train all models in the CM approach are proportional to $K * L$. However, in the ICPTL approach, since the source model is trained per each cluster from scratch and the subsequent models are fine-tuned using $l$ number of epochs ($l << L$), its training time and training resource requirements are proportional to $K*L +(N-K) * l$. Thus, the training time and training resource requirements of the two proposed approaches are greater than that of the GM approach but less than that of the MPD approach. Although the GM approach has the least training resource requirements and also the least model management complexity (since it has only one model), it also has the least accuracy. On the other hand, the MPD approach has the highest accuracy at the cost of the highest training resource requirements. Thus, the proposed clustering-based model training approaches find a good trade-off between the accuracy and efficiency of training resource requirements. Out of the two proposed approaches, the CM approach has higher model management efficiency as it is easier to manage retraining of $K$ cluster models than $N$ individual models per device in the ICPTL approach.
 
\section{Dynamism Handling of Clustering-based Training Approaches}
\label{sec:dynamism}
We proposed the clustering-based training approaches and evaluated them considering the scenario of training models for initial deployment in edge computing environments. However, in dynamic edge computing environments, new edge devices could be added after the first point of deployment, and existing devices could be removed from the infrastructure. Furthermore, the normal data distribution of microservices deployed in edge devices can change due to data drift or osmotic movement of microservices \cite{villari2016osmotic}. This dynamism of edge computing environments poses the requirement to retrain performance anomaly detection ML models. Through this section, we explain how our proposed clustering-based training approaches are capable of handling this requirement while maintaining their initial claims regarding accuracy and efficiency. 

When a new edge device is added to the infrastructure, it will be assigned a cluster. In the ICPTL approach, the model corresponding to that device will retrain from its nearest neighbor. In the CM approach, the existing model of the cluster can be used unless cluster composition is changed.

When an existing edge device is removed from the infrastructure, ICPTL approach does not require any actions. Whereas, in the CM approach, the model can be retrained only if the cluster composition is significantly changed. 

Data drift or osmotic movement can sometimes change the cluster to which an edge device is assigned. In that case, in the ICPTL approach, the model corresponding to the device can retrain from its new nearest neighbor's model. In contrast, the CM approach is required to check for changes in cluster composition of both source and destination clusters and retrain if cluster composition has significantly changed.

\section{Conclusions and Future Work}
\label{sec:conclusion}

Since current performance anomaly detection studies in edge computing environments employ different model training approaches that either prioritize training efficiency or accuracy, this paper introduces two clustering-based model training approaches : (1) intra-cluster parameter transfer learning-based model training (ICPTL) and (2) cluster-level model training (CM) that aims to find a middle ground between the training efficiency and accuracy of anomaly detection models. These methods involve training models on clusters of edge devices that have similar normal data distributions. The ICPTL approach aims to achieve the same accuracy as MPD but with fewer training cycles, by identifying a Minimum Spanning Tree (MST) for each cluster and then training only the device at the beginning of the tree, while the other devices receive the model from the nearest neighbor device and undergo parameter transfer learning to train for a few epochs until they reach acceptable accuracy. On the other hand, the CM approach aims to improve training efficiency even further by reducing the number of models, by training a single model for each cluster, using normal data collected from all edge devices in the cluster. 

The comprehensive and reproducible evaluation conducted on the publicly available and widely-used “Server Machine Dataset (SMD)” revealed that the AutoEncoder (AE) algorithm is the most suitable for anomaly detection due to its high efficiency during inference and the best mean AUC and F1-score compared to other unsupervised approaches. 

Upon evaluating the proposed clustering-based training approaches using the AE algorithm, we could observe that the ICPTL approach achieved the second-highest mean AUC and F1-score while consuming only 40\% of the total training time required by the MPD approach. Furthermore, the CM approach increased training efficiency by consuming 23\% less training time than the ICPTL approach while reducing the total number of models by two-thirds. Additionally, the CM approach demonstrated a 16\% and 30\% improvement in AUC and F1-score, respectively, compared to the GM approach. 

Further experiments conducted on the performance anomaly dataset collected from an emulated edge computing environment reaffirmed the previous results, with the ICPTL approach achieving the accuracy of the MPD approach (which is also the highest accuracy) while requiring only 34\% of the total training time. The CM approach further increased training efficiency by consuming 22\% less training time than the ICPTL approach while reducing the total number of models by two-thirds. In addition, the CM approach showed a 13\% and 5\% improvement in AUC and F1-score, respectively, compared to the GM approach. 

In summary, the ICPTL approach matches the accuracy of the MPD approach with fewer training cycles, and the CM approach further improves training efficiency by reducing the number of models. 

As part of future work, we plan to investigate how to adapt the suggested training approaches to suit various placement strategies apart from QoS and resource requirement-aware placement. Additionally, we aim to explore ways to optimize (especially in terms of efficiency) the subsequent stages of anomaly detection, including anomaly localization and root cause analysis, to better align with the unique features of edge computing environments.

\bibliographystyle{IEEEtran}
	
\bibliography{reference}

\end{document}